\documentclass[aps,prb,twocolumn,groupedaddress]{revtex4-1}
\usepackage{amsmath,amssymb,graphicx,psfrag}
\usepackage{color,epsfig,rotate}
\usepackage{natbib}

\begin{document}

% \preprint{}

\newcommand{\sro}{Sr$_{4}$Ru$_{3}$O$_{10}$}
\newcommand{\srosc}{Sr$_{2}$RuO$_{4}$}
\newcommand{\srocep}{Sr$_{3}$Ru$_{2}$O$_{7}$}
\newcommand{\srofm}{SrRuO$_{3}$}
\newcommand{\sron}{Sr$_{n+1}$Ru$_{n}$O$_{3n+1}$}
%\newcommand{}{}
%\newcommand{}{}
%\newcommand{}{}

\title{Missing magnetism in \sro: Indication for Antisymmetric Exchange Interaction}


\author{Franziska Weickert}
\email[]{weickert@lanl.gov}
\affiliation{MPA-CMMS, Los Alamos National Laboratory, Los Alamos, NM 87544, USA}
\affiliation{NHMFL, Florida State University, Tallahassee, FL 32310, USA}
\author{Leonardo Civale}
\affiliation{MPA-CMMS, Los Alamos National Laboratory, Los Alamos, NM 87544, USA}
\author{Boris Maiorov}
\affiliation{MPA-CMMS, Los Alamos National Laboratory, Los Alamos, NM 87544, USA}
\author{Marcelo Jaime}
\affiliation{MPA-CMMS, Los Alamos National Laboratory, Los Alamos, NM 87544, USA}

\author{Myron B. Salamon}
\affiliation{University of Texas at Dallas, Richardson, TX 75080, Dallas, USA}

%\thanks{}

\author{Emanuela Carleschi}
\affiliation{Department of Physics, University of Johannesburg, Auckland Park 2006, South Africa}
%\author{B. P. Doyle}
%\affiliation{University of Johannesburg, Auckland Park 2006, South Africa}
\author{Andre M. Strydom}
\affiliation{Department of Physics, University of Johannesburg, Auckland Park 2006, South Africa}

\author{R. Fittipaldi}
\affiliation{CNR-SPIN Institute Sede Secondaria di Salerno and University of Salerno, Via Giovanni Paolo II, I-84084 Fisciano, Italy}
\author{V. Granata}
\affiliation{CNR-SPIN Institute Sede Secondaria di Salerno and University of Salerno, Via Giovanni Paolo II, I-84084 Fisciano, Italy}
\author{A. Vecchione}
\affiliation{CNR-SPIN Institute Sede Secondaria di Salerno and University of Salerno, Via Giovanni Paolo II, I-84084 Fisciano, Italy}


\date{\today}

\begin{abstract}
We report a detailed study of the magnetization modulus as a function of temperature and applied magnetic field under varying angle in \sro\ close to the metamagnetic transition at $H_{c}\backsimeq 2.5\,$T for $H \perp c$. We confirm that the double-feature at $H_{c}$ is robust without further splitting for temperatures below 1.8\,K down to 0.48\,K. The metamagnetism in \sro\ is accompanied by a reduction of the magnetic moment in the plane of rotation and large field-hysteretic behavior. The double anomaly shifts to higher fields by rotating the field from $H\,\perp \,c$ to $H\,\parallel\,c$.
We compare our experimental findings with numerical simulations based on spin reorientation models caused by intrinsic magnetocrystalline anisotropy and Zeeman effect. Crystal anisotropy is able to explain a metamagnetic transition in the ferromagnetic ordered system \sro, but a Dzyaloshinskii-Moriya term is crucial to account for a reduction of the magnetic moment as discovered in the experiments.
\end{abstract}


\maketitle

%\subsection{}
%\subsubsection{}

\section{\label{intro}Introduction}

\sro\ belongs to the Ruddlesden-Popper family of ruthenium oxide perovskites \sron. This class of metallic compounds caught much attention in recent years due to its rich variety of ground states. \srosc\, the $n=1$ member, is discussed as an example of rare $p$-wave superconductivity.\cite{ishida_98} A quantum critical endpoint covered by a high entropy phase was found in the $n=2$ layer system \srocep.\cite{grigera_01a} The compound \sro\ ($n=3$) discussed here shows ferromagnetism below $T_{C}=105\,K$.\cite{crawford_02} Neutron diffraction experiments in zero magnetic field reveal ordering of the Ru moments along the $c$-axis. No ferromagnetic (FM) or antiferromagnetic (AFM) correlations are observed in the $ab$-plane.\cite{granata_13,granata_16} \sro\ contains four inequivalent Ru sites with two different magnetic moments of 0.9\,$\mu_{B}$ and 1.5\,$\mu_{B}$ sitting on outer and inner RuO layers, respectively. The magnetic unit cell contains 8 of the smaller and 4 of the bigger magnetic moments averaging to 1.1\,$\mu_{B}$ per Ru.
The higher order ruthenate SrRuO$_{3}$ with $n=\infty$ orders also FM at a Curie temperature of 165\,K.\cite{callaghan_66,kanbayasi_76}. \sron\ are strongly 2-dimensional electron systems with the trend to become more isotropic for higher $n$, because of their layered structure. Two-dimensionality is reflected in anisotropic transport properties as seen for \sro\ in the ratio of the electrical resistivity $\rho_{c}/\rho_{ab} \simeq 400$\cite{fobes_10} and confirmed by optical conductivity experiments.\cite{mirri_12}

Metamagnetism is a phenomenon observed in magnetic materials, where hidden magnetism is suddenly uncovered by the application of an external field. The origin of metamagnetism ranges from spin flip transitions in antiferromagnets,\cite{held_97,mun_16} to the reconstruction of the Fermi surface in metallic materials,\cite{deppe_12} and in the vicinity of a quantum critical point.\cite{weickert_10} In a general description, metamagnetism is a phase transition or crossover from a magnetically disordered or ordered state with small net magnetization to a field polarized (FP) or partially FP state. In the case of \sro\, the term metamagnetism refers to the sudden increase in the magnetization when the field applied in the ($ab$) in-plane of this layered compound exceeds ~2\,T. Magnetism is hidden only because the spontaneous moment is mainly aligned with the easy $c$-axis at smaller fields; we continue to refer to this as metamagnetic (MM) transition.
%Crucial in the description of metamagnetism are symmetry considerations.
%The here investigated compound \sro\ holds an exceptional position, because it develops metamagnetic (MM) anomalies deep %inside the ferromagnetic (FM) phase at $2.5\,$T where FM domain reorientation processes are fully completed.

While metamagnetism of $H \perp c$ in \sro\ was discovered from early on in flux grown single crystals,\cite{crawford_02} it took more than a decade to improve the crystal quality to a level to see a double step in the magnetization at the MM transition.\cite{carleschi_14} This strong dependence of physical properties on the crystal purity is a characteristic signature of strontium ruthenates \sron\ and was also observed in the sister compound \srocep.\cite{grigera_01a,perry_04c,grigera_04,gegenwart_06a} The MM transition in \sro\ develops below 68\,K as a double-transition close to zero field and shifts gradually to $H_{c}\sim2.5\,$T with temperatures down to 1.7\,K.\cite{carleschi_14} \emph{Carleschi et al.}\cite{carleschi_14} speculate that the double transition originates either in the ordering of Ru magnetic moments on two inequivalent crystallographic sites or in the presence of two van Hove singularities in the density of states close to the Fermi level. A transport study based on electrical resistivity\cite{mao_06,fobes_07} reveals steps in the magnetoresistance at various critical fields around $H_{c}$ accompanied by pronounced hysteresis.
%Interestingly, the critical fields where the steps occur change for measurements done at different sample cool-downs and are %only visible in measurements with decreasing magnetic field.
\emph{Fobes et al.}\cite{fobes_07} interpret the transport data as domain movement of regions with high and low electronic spin polarization. Anomalous behavior at the MM transition was also observed in specific heat experiments up to 9\,T\cite{cao_07} and in thermopower investigations.\cite{xu_07} Neutron diffraction experiments up to 6\,T reveal a change of lattice parameters at the critical field $H_{c}$.\cite{granata_13} Field and pressure dependent Raman measurements\cite{gupta_06} as well as a recent study of thermal expansion and magnetostriction\cite{schottenhamel_16} confirm strong magnetoelastic coupling in \sro.

%Investigations of the angle dependence of the MM transition by torque magnetometry at 4.2\,K report the bifurcation of the metamagnetic anomaly and %the shift to higher and lower fields, respectively.\cite{jo_07}

In this work we focus on magnetization measurements up to 7\,T and down to lowest temperatures of 0.46\,K and under rotational fields between the $c$-axis and the $ab$-plane as well as ($ab$) in-plane rotation. Our investigations include a detailed analysis of the behavior of the magnetization modulus $M$ at the MM transition and its individual components $M_{ab}$ and $M_{c}$ simultaneously. In the following, we analyze and interpret our data in a localized picture, meaning the magnetic moments are mainly confined on the Ru$^{4+}$ sites in the crystal structure of \sro. This scenario is supported by neutron diffraction experiments which have determined the spin and orbital momentum distribution in great detail.\cite{granata_13,granata_16} Our main discovery is the observation of a reduced measured moment at the MM transition caused by a spin component pointing out of the rotational plane which we assert can best be explained by significant anisotropic exchange interactions in \sro.
%There are several aspects in \sro\ that aren't well understood and we would like to address in this article. What is the %microscopic mechanism of the MM transition? Does it split further for temperatures below 1.8\,K? Does the MM transition %enclose a true thermodynamic phase with broken symmetry? Does it resemble similarities with the nematic order observed in %\srocep ? Most importantly, what is the origin of the double-step?
%Since torque magnetometry can be imprecise in strongly anisotropic materials, we were motivated to investigate at small %angles and down to lower %temperatures.

\section{\label{res}Results}

\begin{figure}
	\includegraphics[width=3in]{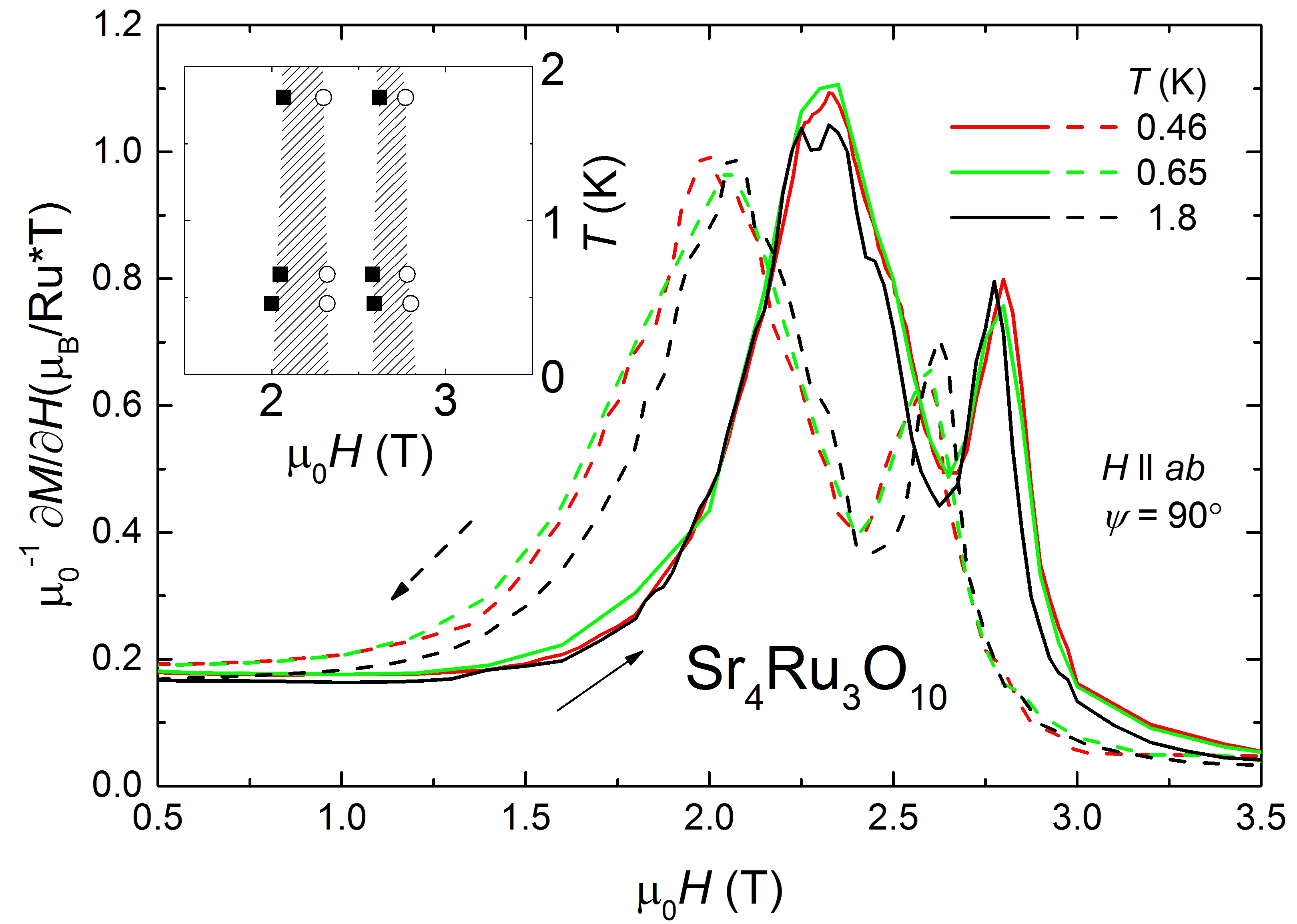}
	\caption{\label{dMdHvsH}Field derivative of the magnetization $\partial M / \partial H$ at 1.8\,K, 0.65\,K and 0.46\,K. Solid lines show measurements during increasing field sweeps and dashed lines during decreasing field sweeps as labeled by arrows. The inset shows $H-T$ phase diagram close to the MM transition with regions of hysteresis marked as striped patterns.}
\end{figure}
At first, we focus on the magnetization measured for $H \perp c$ at temperatures below 2\,K. Figure \ref{dMdHvsH} shows $\chi_{AC}=\partial M/\partial H$ between 0.5\,T and 3.5\,T. We observe a clear double MM phase transition with a main anomaly at $H_{c1}=2.3$\,T and a second anomaly at $H_{c2}=2.8$\,T for increasing field as observed by \emph{Carleschi et al.}.\cite{carleschi_14} Our new experimental data down to 0.46\,K clarify that neither transition sharpens to lower temperatures nor is there splitting into more distinct anomalies. 
%The scattering of the data points is in the experimental error bars. 
The inset in Fig. \ref{dMdHvsH} shows the $H-T$ phase diagram with near-vertical phase boundaries for $T \rightarrow 0$ at the MM transition. Both anomalies are shifted by $-0.3$\,T for measurements in decreasing magnetic field. The size of the hysteretic region, marked as striped pattern, remains similar for all temperatures below 1.8\,K.

%\begin{figure}
%	\includegraphics[width=2in]{figures/sketch_small.jpg}
%	\caption{\label{sketch} Sketch of the sample geometry of \sro\ mounted inside the SQUID magnetometer. $\psi$ is the rotation angle of the applied %field $\vec{H}$ and $\theta$ the angle of the magnetization $\vec{M}$, both in respect to the magnetic easy $c$-axis. $M_{long}$ and $M_{trans}$ are %measured components of $\vec{M}$ parallel and perpendicular to the applied field in the plane of rotation. The component $M_{perp}$ parallel to the axis %of rotation is not captured during the measurement.}
%\end{figure}
\begin{figure}
	\includegraphics[width=3in]{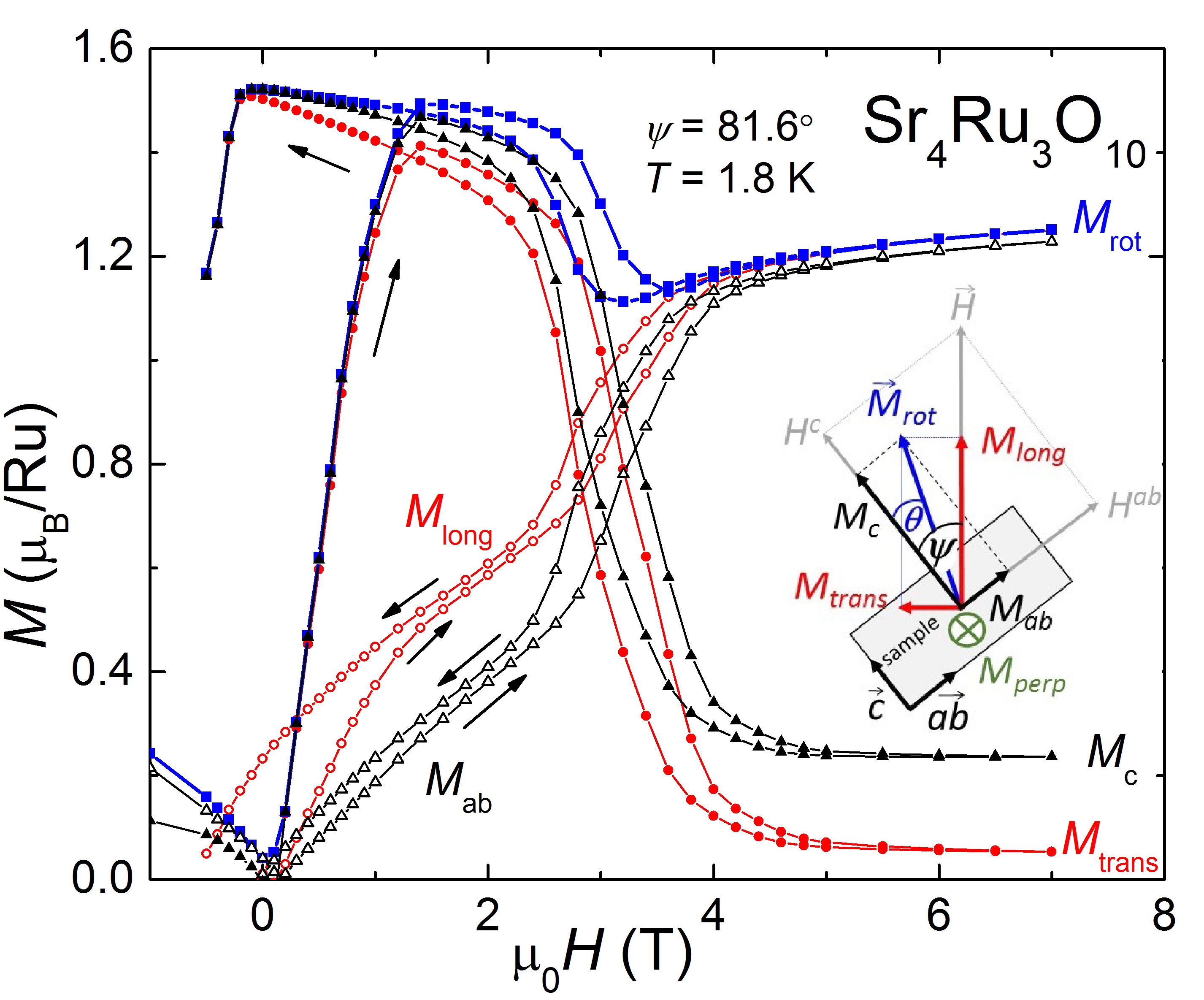}
	\caption{\label{Mexample}The inset shows a geometrical sketch of the sample \sro\ mounted inside the SQUID magnetometer. $\psi$ is the rotation angle of the applied field $\vec{H}$ and $\theta$ the angle of the magnetization $\vec{M}$, both in respect to the magnetic easy $c$-axis. $M_{long}$ and $M_{trans}$ are measured components of $\vec{M}$ parallel and perpendicular to the applied field in the plane of rotation. The component $M_{perp}$ parallel to the axis of rotation is not captured during the measurement. The main panels compares the different components of the magnetization $M_{long}$, $M_{trans}$, $M_{ab}$, $M_{c}$, and the modulus $M_{rot}$ versus magnetic field $H$ measured at 1.8\,K for $\psi = 81.6^{\circ}$.}
\end{figure}
The operation mode of the SQUID magnetometer allows the simultaneous collection of longitudinal $M_{long}$ and transversal component $M_{trans}$ of the sample magnetization in respect to the applied magnetic field $\vec{H}$ as sketched in the inset of Fig.\,\ref{Mexample}. The magnetization component $M_{perp}$ occurring perpendicular to the rotational plane is not recorded during the measurements. This geometry allows us to calculate the magnetization $M^{2}_{rot} = M^{2}_{long} + M^{2}_{trans}$ in the rotational plane. The knowledge of the rotation angle $\psi$ and relation $\tan(\psi - \theta) = M_{trans}/M_{long}$ enables the determination of the angle of the magnetization $\theta$ with respect to the magnetic easy axis $c$ in the rotational plane. We can now estimate $M_{ab} = M_{rot} \sin \theta$ and $M_{c} = M_{rot} \cos \psi$ magnetization, again occurring in the plane of rotation. Note, we follow closely the notation of angles used for magnetic anisotropic materials.  Figure \ref{Mexample} illustrates the different components of the magnetization for one particular measurement with $\psi = 81.6^{\circ}$ taken at 1.8\,K. Prominent feature is the hysteresis loop around $\pm 1$\,T, caused by FM domain dynamics. The longitudinal magnetization $M_{long}$ increases moderately in small fields and shows a sudden rise at $H_{c}\backsimeq3$\,T at the MM transition. $M_{trans}$ on the other hand consists mainly of the $M_{c}$ component with a sudden decrease of the magnetization at the same critical field $H_{c}$. $M_{trans}\neq 0$ above $H_{c}$ indicates incomplete field polarization meaning that $\vec{M}$ is not perfectly aligned with $\vec{H}$. This observation points to the presence of magnetic anisotropy. The calculated magnetization modulus $M_{rot}$ in the plane of rotation is depicted as black line in Fig.\,\ref{Mexample}. We find a maximum moment of $1.5\,\mu_{B}$ slightly higher than obtained in neutron experiments,\cite{granata_13,granata_16} but in good agreement with previous magnetization studies.\cite{cao_03,zhou_05} Most peculiar is that $M_{rot}$ drops suddenly below 1.2\,$\mu_{B}$ at the MM transition and only recovers partially to 1.2\,$\mu_{B}$ up to maximum applied field of 7\,T. This missing component of the magnetic moment in \sro\ was never recognized before. Furthermore, we observe strong hysteresis at the MM transition between up and down measurements as reported in previous investigations.\cite{cao_03, zhou_05,fobes_07,fobes_10,carleschi_14}

\begin{figure}
	\includegraphics[width=3in]{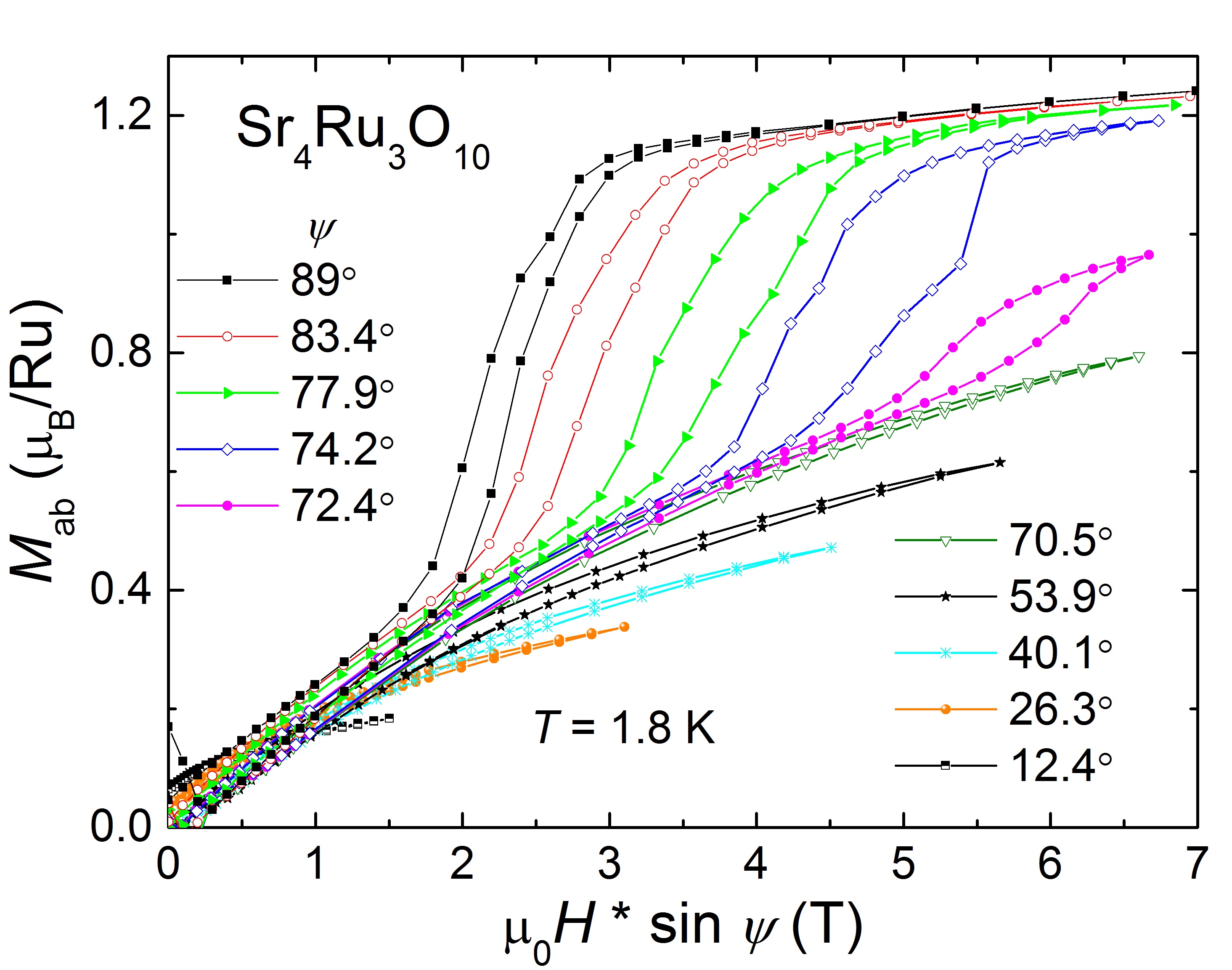}
	\caption{\label{MabvsHsin}$ab$-plane magnetization $M_{ab}$ as a function of $H^{ab} = H \sin \psi$ is shown for angles $\psi$ between 89$^{\circ}$ and 12.6$^{\circ}$ measured at 1.8\,K.}
\end{figure}
\begin{figure}
	\includegraphics[width=3.5in]{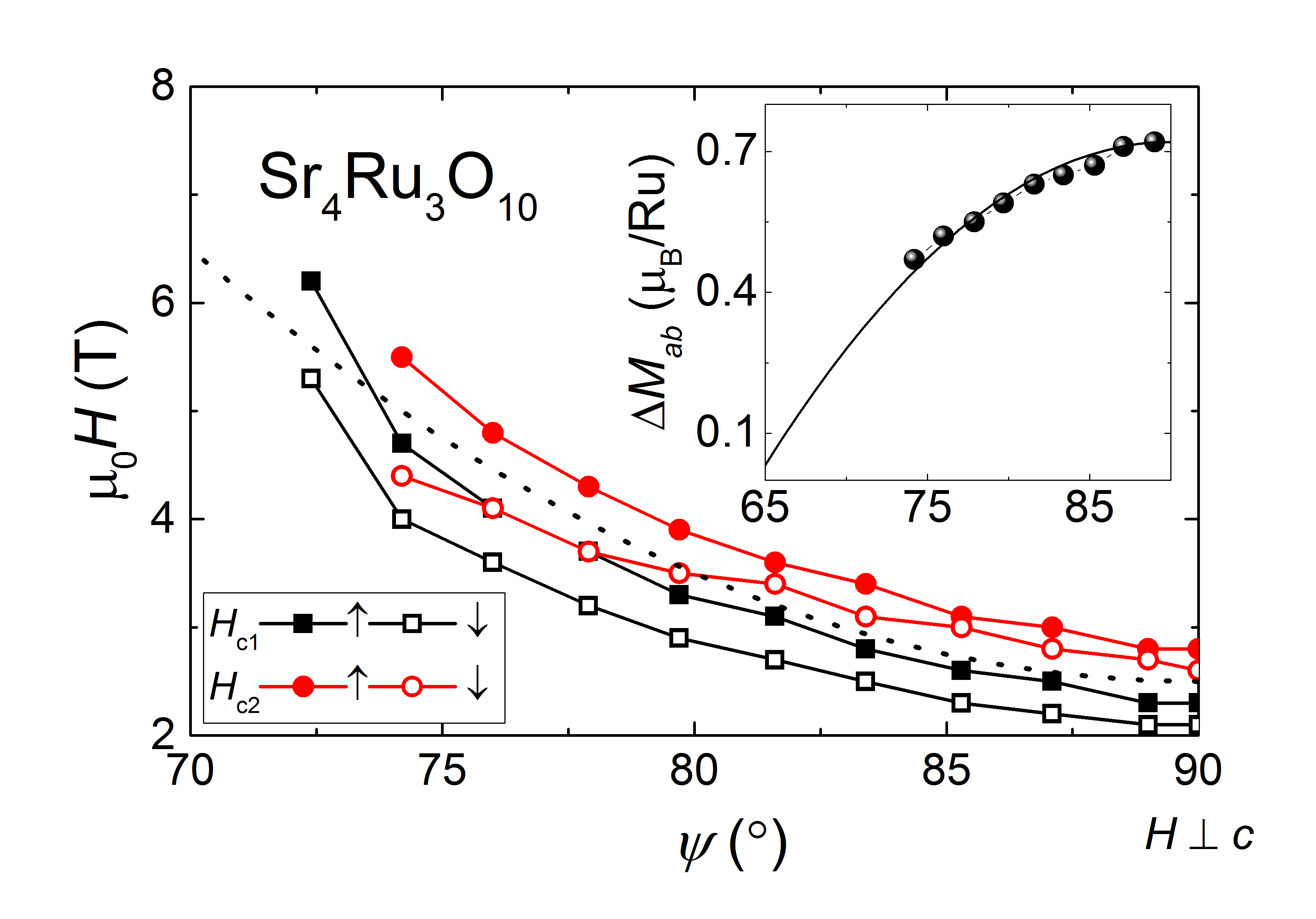}
	\caption{\label{phasedia}Angular $\psi$ dependent shift of the double anomaly at the MM transition in \sro. Solid points mark positions for increasing and empty points decreasing field sweeps. The dotted line is a quadratic fit to the data. The inset shows the reduction of the magnetization step $\Delta M_{ab}$ at the MM transition as a function of $\psi$ including a quadratic extrapolation marked as solid line.}
\end{figure}
Geometrical effects can distort magnetic properties during magnetization experiments. Therefore, we plot in Fig.\,\ref{MabvsHsin} $M_{ab}$ as a function of the field component in the $ab$-plane $H^{ab}=H \sin \psi$ to examine how $H_{c1,2}$ change with $\psi$. In contrast to previous results by \textit{Jo et al.}\cite{Jo_07} obtained by torque magnetometry, we observe a clear simultaneous increase of both critical fields $H_{c1,2}$ to higher values while rotating from $H \perp c$ to $H \parallel c$. In fact, $H_{c1,2}$ move out of the observable field range of 7\,T maximum for $\psi \lesssim 72^{\circ}$. Fig.\,\ref{phasedia} summarizes the critical fields $H_{c1,2}$ in the $H-\psi$ phase diagram for field up and down sweep measurements. The difference $H_{c1}-H_{c2}$ increases slightly with smaller $\psi$. As mentioned above, the double anomaly is accompanied by significant hysteresis. The inset of Fig.\,\ref{phasedia} shows the evolution of combined step size $\Delta M_{ab}$ of both MM transitions for decreasing $\psi$. It follows a quadratic fit function marked as solid line and extrapolates to zero at about $65^{\circ}$.

\begin{figure}
\includegraphics[width=3in]{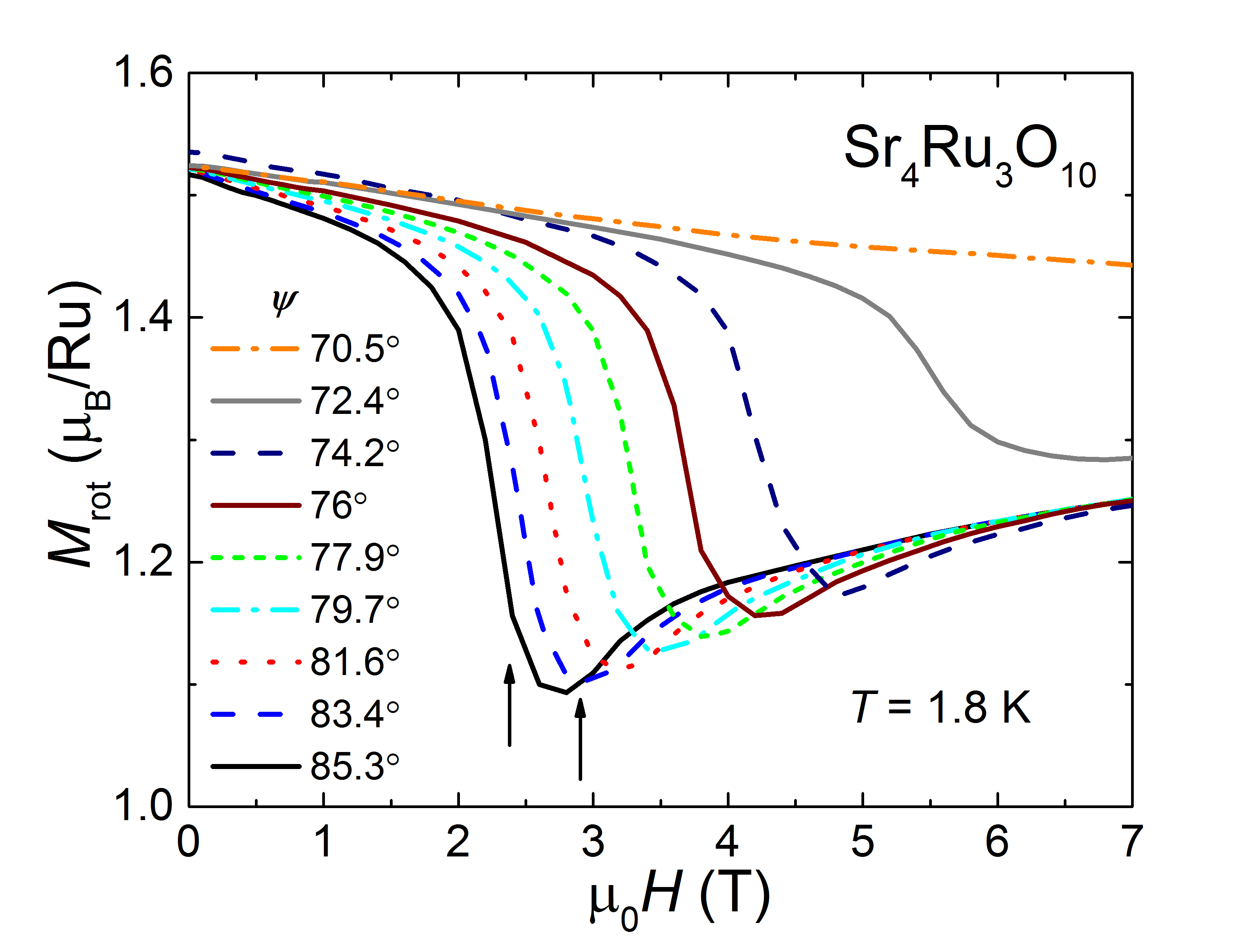}
\caption{\label{modMvsH}Magnetization modulus $M_{rot}$ in the rotational plane versus magnetic field $H$ for angles $\psi$ between 85.3$^{\circ}$ and 70.5$^{\circ}$ in decreasing fields. Arrows mark anomalies at $H_{c1}$ and $H_{c2}$ for the measurement at $\psi = 85.3^{\circ}$.}
\end{figure}
The magnetization modulus $M_{rot}$ recorded in the plane of rotation for $\psi$ between 85.3$^{\circ}$ and 70.5$^{\circ}$ is plotted in Fig.\,\ref{modMvsH}. We only show field-down sweep measurements for clarity. Striking is the occurrence of a drop from about 1.5\,$\mu_{B}$ to below 1.2\,$\mu_{B}$ at the critical field of the main anomaly $H_{c1}$ followed by a minimum and a small step at the second anomaly at $H_{c2}$. The described features are marked in Fig.\,\ref{modMvsH} by arrows for the measurement at $\psi = 85.3^{\circ}$. The MM anomaly broadens and moves to higher fields for decreasing angles $\psi$.

\section{\label{disc}Discussion}
The "loss" of magnetic moment in the rotational plane can be explained either by partial AFM alignment or by a moment $M_{perp}$ occurring perpendicular to the rotation (parallel to rotation axis of $\psi$). First scenario can be excluded based on neutron experiments where no short or long range AFM coupling neither in zero nor in magnetic fields $H > H_{c}$  was observed in the $ab$-plane.\cite{granata_13,granata_16} The second scenario is rather unexpected since magnetic moments tend to align with field and stay within the rotational plane, if no further coupling is present. We want to focus in our discussion on two mechanisms that potentially lead to a $M_{perp}$ component in the magnetization. First one is based on general magnetocrystalline anisotropy in tetragonal symmetry, with an easy $c$-axis and 4-fold in-plane anisotropy. The second mechanism is antisymmetric exchange between spins, also called Dzyaloshinskii-Moriya (DM) interaction, causing a canting of the spins $\vec{S_{i}}\times \vec{S_{j}}$.

\subsection{General Anisotropy}

We have to have a closer look at the crystal structure of \sro\ in order to understand and model its magnetic anisotropy caused by spin-orbit coupling. \sro\ crystals consist of three layers of corner sharing RuO$_{6}$ octahedra separated by a double layer of Sr-O. Primary Bragg reflections in synchrotron experiments can be indexed assuming a tetragonal unit cell with space-group $I4/mmm$, but a more detailed analysis of secondary reflections reveals orthorhombic $Pbam$ symmetry.\cite{crawford_02} The lower symmetry originates in $c$-axis rotation of the RuO$_{6}$ octahedra that are correlated between different layers, meaning $+11.2^{\circ}$ clockwise rotation for inner and $-5.6^{\circ}$ counterclockwise rotation for outer layers.

The free energy $F$ accounting for magnetocrystalline anisotropy in a tetragonal lattice, can be modeled\cite{horner_68} by
\begin{equation}\label{equ_general}
       F = F_{0}+K_{1} \cos^{2}\theta +(K_{2}+K_{3}\sin(2\varphi))\sin^{4}\theta - F_{Z}.
\end{equation}
$F_{0}$ is a constant background contribution independent of $\vec{H}$ or $\vec{M}$. $\vec{M}$ is expressed in polar coordinates ($\theta, \varphi$), with $\theta$ = 0 along the crystal $c$-axis and $\varphi$ = 0 defining the in-plane hard axis for $K_{3} < 0$. Here, $K_{1} < 0$ defines the easy direction and the Zeeman term $F_{Z}$ can be written as
\begin{equation}\label{equ_zeeman} 
	 F_{Z} = M H (\sin \theta \sin \psi (\cos \omega \cos \varphi + \sin \omega \sin \varphi) + \cos \theta \cos \psi).
\end{equation}
The applied field has the polar coordinates ($\psi, \omega$), with $\omega = 0$ corresponding to field rotation from the $c$- axis to the in-plane hard direction and $\omega = \pi /4$, field rotation to the easy direction.

We used numerical minimization of equation (\ref{equ_general}) to determine $\theta$ and $\varphi$ as functions of the applied field. In the uniaxial case ($K_{3}=0$), the MM behavior in $M_{ab}$ and $M_{c}$ at the critical field $H_{c}$ is reproduced by choosing correct parameters $K_{1}$ and $K_{2}$ (data not shown). However, uniaxial anisotropy is unable to reproduce any reduction of the magnetization in $M_{rot}$, since the moment always stays in the rotational plane.  

Note, we do not know precisely the in-plane orientation of our sample. However, the rectangular shape suggests that $\psi$ rotation axis is parallel to one of the principal axes  such as [100] or [110]. For 4-fold tetragonal symmetry either one of them would be the intermediate or hard axis, respectively. We consider in the following a projection of $\vec{H}$ onto the magnetic hard axis in the $ab$-plane with $\omega = 0$, because tilting of $\vec{M}$ towards the hard axis forces the magnetic moments to align spontaneously toward either one of the intermediate axes, which are $45^{\circ}$ apart from the hard axis. This spontaneous alignment $\pm 45^{\circ}$ is energetically degenerated and could lead to domain formation with an overall smaller net magnetization as observed in our measurements.

\begin{figure}
	\includegraphics[width=3in]{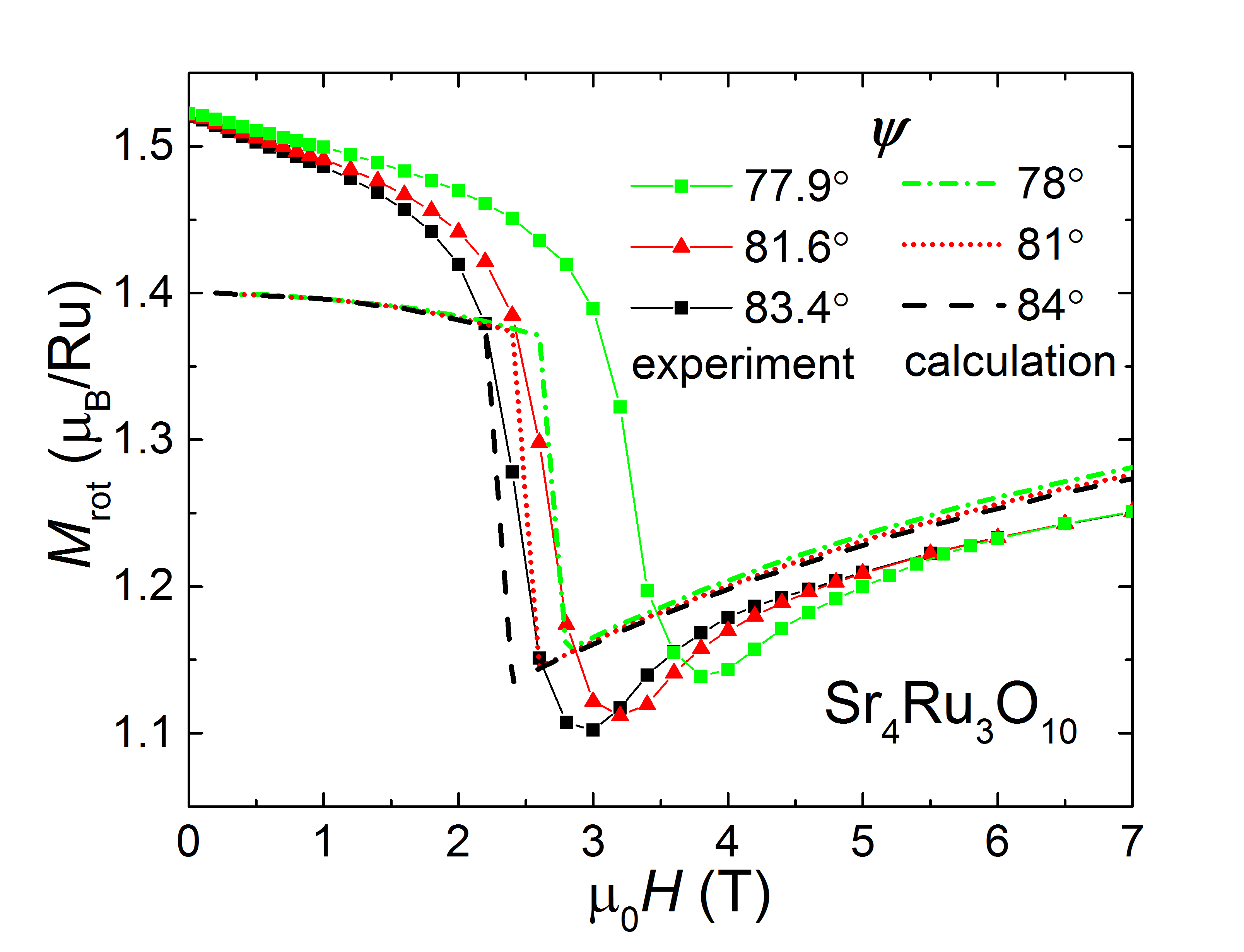}
	\caption{\label{modelMvsH} The experimental magnetization moduli $M_{rot}$ for 3 different angles $\psi$ (dots) measured at 1.8\,K in decreasing magnetic field are compared with numerical simulations (lines) of the general anisotropy model as described by Eq.\,(\ref{equ_general}).}
\end{figure}
Fig.\,\ref{modelMvsH} compares numerical results of $\psi=78^{\circ}, 81^{\circ}, 84^{\circ}$ based on equation (\ref{equ_general}) with experimental data of the magnetization modulus $M$ for $\psi = 77.9^{\circ}, 81.6^{\circ}, 83.4^{\circ}$. We are able to reproduce i) a critical field value $H_{c}\backsimeq 2.5$\,T that increases with smaller $\psi$, ii) a drop $\Delta M$ at $H_{c}$ that is comparable in size with the experimental data, and iii) a gradual slope $M(H)$ for $H>H_{c}$. The double feature at the MM transition is missing due to the simplicity of the model. We obtain anisotropy parameters $K_{1}=3.1$\,K, $K_{2}= 0.1$\,K and $K_{3}= -2.2$\,K in Kelvin energy scale which convert to the following values 300\,kJ/m$^{3}$, 10\,kJ/m$^{3}$, and -210\,kJ/m$^{3}$, respectively, in units widely used in magnetic anisotropy tables. The 4th order parameter $K_{2}$ being more than 10 times smaller than $K_{1}$ implies that it is irrelevant for the description of the anisotropy in \sro. For comparison, the $3d$ FM metal cobalt has anisotropy constants of $K_{1}= 450$\,kJ/m$^{3}$ and $K_{2}= 150$\,kJ/m$^{3}$, which are of similar size as $K_{1}$ in \sro .\cite{cullity_08}

\begin{figure}
	\includegraphics[width=3in]{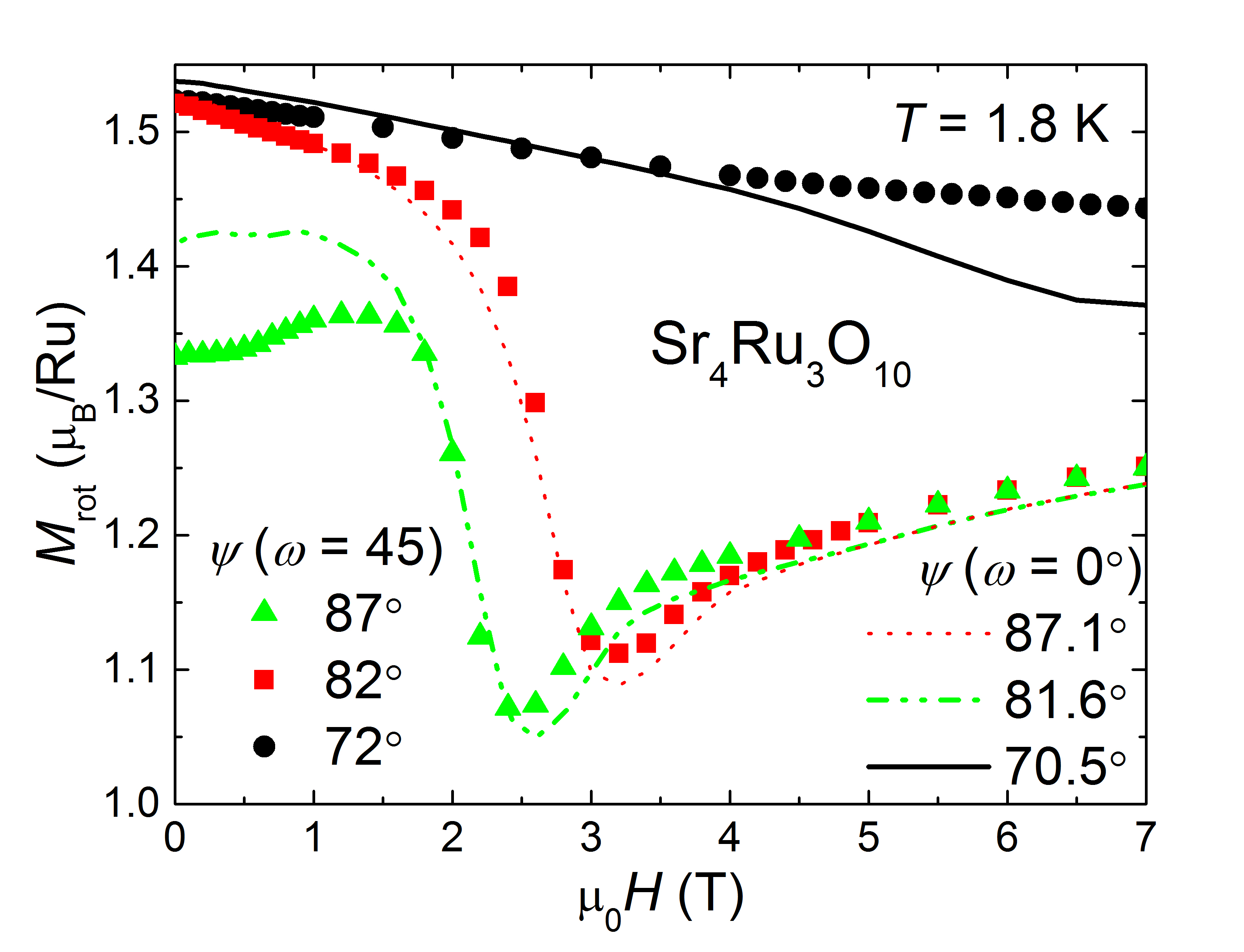}
	\caption{\label{45deg}Decreasing field $H$ measurements of the magnetization modulus $M$ taken at 1.8\,T at three angles $\psi \backsimeq 87^{\circ}, 82^{\circ}$, and $71^{\circ}$ are shown for two different in-plane angles $\omega = 0$ and $45^{\circ}$. The almost identical results for $0$ and $45^{\circ}$ disable magnetocrystalline anisotropy being the cause for the loss of magnetic moment in \sro.}
\end{figure}
Despite the reasonable agreement between experiment and model, it is necessary to check in a subsequent experiment our initial assumption of tilting the spins towards the magnetic hard axis in the $ab$-plane. Therefore, we rotate the sample by $\omega = 45^{\circ}$ in the plane and measure again $M$ at three different angles $\psi$ as shown in Fig.\,\ref{45deg}. We anticipate the 45$^{\circ}$ change would bring the intermediate anisotropy axis into the rotation plane. Specifically, the magnetization would rotate toward the intermediate axis with the total magnetization remaining in the rotation plane and therefore no "loss" of magnetic moment effect. Surprisingly, the magnetization $M_{rot}$ shows exactly the same behavior as for the $\omega = 0$ experiments within experimental uncertainty. Even if in both experiments $\omega = 0$ and 45$^{\circ}$, the plane of $\psi$-rotation would not include exactly the principal axes, we would at least expect the observation of a reduced anomaly in $M_{rot}$ at $H_{c}$. Based on our last finding, we exclude general magnetic anisotropy as sole cause for the reduction of moment at the MM transition in \sro.

\subsection{Antisymmetric Exchange Coupling}

\begin{figure}
	\includegraphics[width=3in]{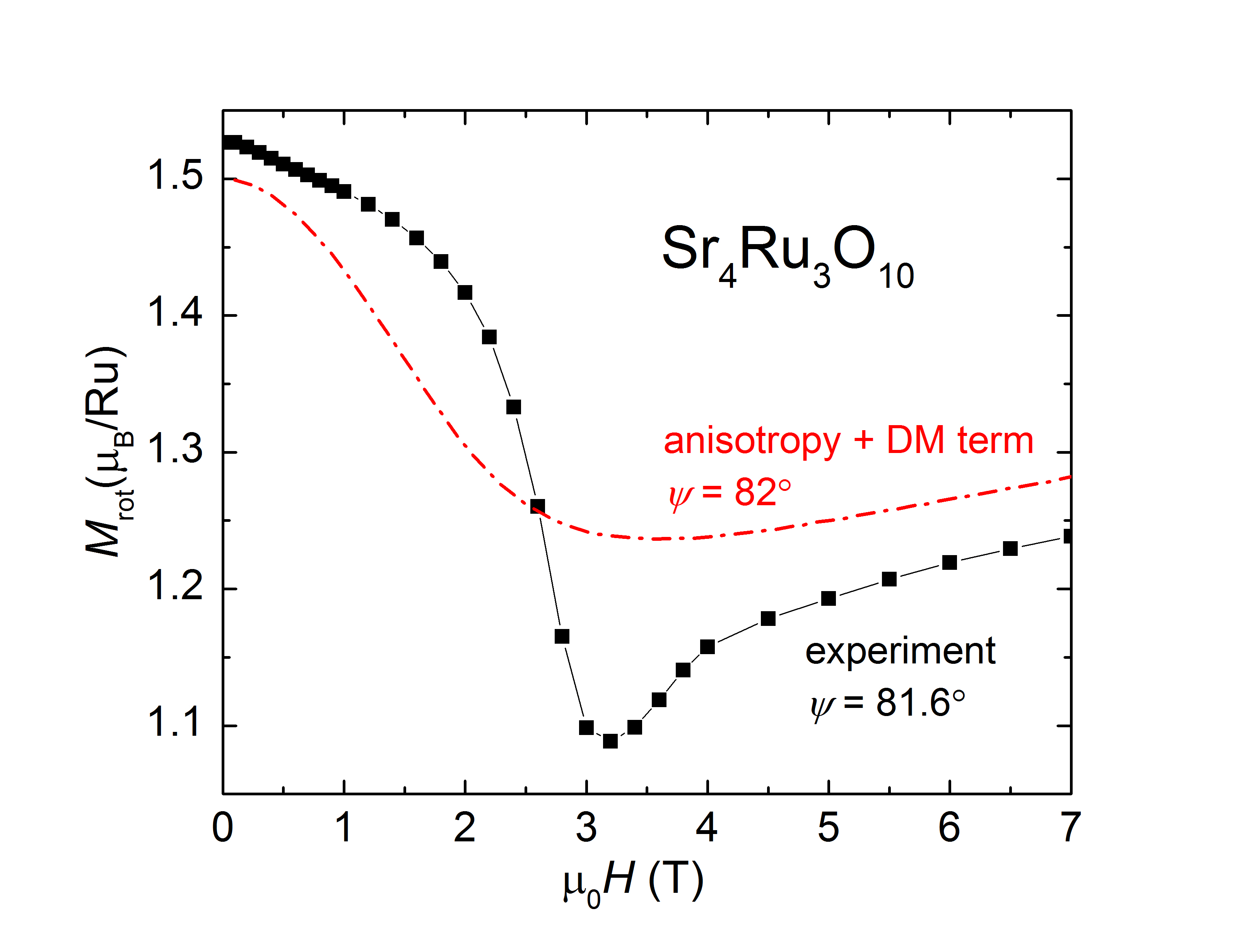}
	\caption{\label{dm}Experimental magnetization modulus $M_{rot}$ in the plane of rotation as a function of magnetic field $\mu_{0}H$ up to 7\,T ($\blacksquare$) and numerical results including tetragonal magnetocrystalline anisotropy and a DM alike energy term (broken line) for $\psi \backsimeq 82^{\circ}$.}
\end{figure}
Anisotropic exchange interactions caused by spin-orbit coupling under certain symmetry constraints were first considered by \textit{Dzyaloshinsky and Moriya}\cite{dzyaloshinsky_58,moriya_60} to explain weak FM ordering inside an AFM phase in transition metal oxides. Recently, \textit{Bellaiche et al.}\cite{bellaiche_12} pointed out that tilting of oxygen octahedra in perovskites can be described by a pseudo-vector $\nu_{i}$ at spin $S_{i}$ position $i$ that leads to an energy reduction 
\begin{equation}\label{equ_bell}
    \Delta E = K \sum_{i,j} (\nu_{i} - \nu_{j}) \cdot (\vec{S}_{i}\times \vec{S}_{j})
\end{equation}
in analogy to DM antisymmetric exchange coupling. The summation is done over nearest-neighboring spins $\vec{S}_{i,j}$ and $K$ is a constant. Consequently, we approximate a DM interaction between the in-plane component of the Ru center magnetization and that of the two nearest-neighbor Ru atoms, whose spins are assumed to remain parallel. This gives rise to an effective energy term
\begin{equation}\label{equ_dm}
F_{DM} = -D \sin^{2}(\theta)\sin(2\varphi)
\end{equation}
in replacement of the $K_3$ term of the magnetocrystalline anisotropy. Angle $\varphi$ is interpreted as the angle between in-plane magnetic moments sitting on one inner and two outer layers with $\varphi = 0$ being the direction of in-plane magnetic field. The change in the dependence on angle $\varphi$ between the magnetization vector and the $c$-axis prevents an accurate minimization of the total energy. Nonetheless, the approximate solution shown in Fig.\,\ref{dm} for $\psi = 82^{\circ}$ does produce a "missing" portion of the total magnetization, with the minimum moving to higher field with decreasing $\psi$. The parameter $D$ is approximately 5.3\,K and comparable to the energy scale of the magnetocrystalline anisotropy $K_1$.

We want to point out that the opening of $2\varphi$ between the inner and outer magnetic moments can be seen as AFM order if it occurs periodically along the $c$-axis. Neutron scattering experiments on holmium-yttrium superlattices\cite{fuente_99} e.g., were able to distinguish between different types of AFM periodicity along $[00l]$ with increasing in-plane magnetic field, such as helical, helifan-shaped, fan-shaped and FM order. However, the particular crystal structure of \sro\ with three layers of RuO$_{6}$ octahedra connected through a double layer of Sr-O doesn't give rise to additional periodicity, even if the magnetic moments follow a repetitive pattern such as $-\varphi, +\varphi, -\varphi$ along $c$ inside the triple layer. Consistently, in-plane AFM ordering was excluded by neutron scattering studies.\cite{granata_13}

% \begin{figure}
% \end{figure}

\section{\label{sum}Summary}

In summary, the MM transition in \sro\ as a function of magnetic field and rotation angle between $H \perp c$ and $H \parallel c$ has been studied in great detail by magnetization measurements down to lowest temperatures in a SQUID magnetometer. Our experimental results reveal a reduced magnetic moment in the plane of rotation which was never recognized before. It is robust to ($ab$) in-plane rotation. We find furthermore that the double step at the MM transition is stable down to lowest temperatures of 0.46\,K. Our experimental results are interpreted in a strict localized picture with magnetic moments of different sizes sitting on inner and outer RuO$_{6}$ layers in the crystal structure. We completed our study with numerical calculations based on energy minimization including Zeeman effect, magnetocrystalline anisotropy and antisymmetric exchange and compared them with the experimental data. We conclude that all three contributions are essential ingredients to understand the magnetization behavior and establish that a \textit{Dzyaloshinsky and Moriya} like component is essential to understand the occurrence of a reduced magnetic moment in \sro.

\section{\label{exp}Methods}

\sro\ single crystals were grown in an image furnace by a floating zone technique\cite{fittipaldi_07} and characterized by energy dispersive spectroscopy, scanning electron microscopy, electron backscattering diffraction and x-ray diffraction techniques. 

%We used the very same single crystal that revealed the double metamagnetic transition for the first time.\cite{carleschi_14}
Magnetization measurements in fields up to 7\,T were carried out in a Quantum Design MPMS SQUID magnetometer equipped with a standard $^{4}$He setup for measurements between 1.7\,K and 100\,K and in an iQuantum $^{3}$He insert that fits inside the MPMS sample space for the temperature range 0.46\,K to 2\,K. Excellent agreement between the $^{3}$He and $^{4}$He data was observed in the temperature range of overlap. Angular dependent measurements at 1.8\,K were obtained with a mechanical rotator mounted inside the MPMS magnetometer. Note that the MPMS operates with a pair of coils for signal detection mounted parallel (longitudinal coil) and perpendicular (transversal coil) to the applied magnetic field. The rotator is aligned with the rotation axis normal to the plane defined by both SQUID coil axes.

The measured single crystal has a rectangular shape of (1.87 x 2.99)mm$^{2}$ in $ab$ and 0.54\,mm along the crystallographic $c$-direction. We considered demagnetization effects by approximating the sample shape with an ellipsoid\cite{osborn_45} and estimated small correction fields of -\,30 mT for $H \perp c$ and -\,140 mT for $H \parallel c$.

\begin{acknowledgments}
Work at Los Alamos was supported by the National Science Foundation, Division of Material Research under Grant NSF-DMR-1157490 and the U.S. Department of Energy, Office of Science, Basic Energy Sciences, Materials Sciences and Engineering Division. 
\end{acknowledgments}

\bibliography{article_201701}{}
% You should use BibTeX and apsrev.bst for references
% Choosing a journal automatically selects the correct APS
% BibTeX style file (bst file), so only uncomment the line
% below if necessary.
\bibliographystyle{apsrev4-1}
\end{document}